\documentclass{aa}  

\usepackage{graphicx}
\usepackage{txfonts}
\usepackage{xspace}
\usepackage{natbib}
\usepackage{color}
\usepackage{textcomp}
\xspace
\newcommand{\veso}{V\#641}\xspace
\newcommand{\vpol}{V-pol}\xspace

\begin{document}

   \title{An independent distance estimate to the AGB star R~Sculptoris}

%  \subtitle{Phase lag measurements }

  \author{M.~Maercker
          \inst{1}
          \and
          M.~Brunner \inst{2}
          \and
          M.~Mecina \inst{2}
          \and
          E.~De~Beck \inst{1}
         %\fnmsep\thanks{Just to show the usage of the elements in the author field}
          }

     \institute{Department of Space, Earth and Environment, Chalmers University of Technology, Onsala Space Observatory, 43992 Onsala, Sweden\\
   \email{maercker@chalmers.se}
         \and
             Department of Astrophysics, University of Vienna, T\"urkenschanzstr. 17, 1180 Vienna, Austria\\
             %\thanks{The university of heaven temporarily does not accept e-mails}
             }

  \date{Received 6 October 2017 / Accepted 22 November 2017}

% \abstract{}{}{}{}{} 
% 5 {} token are mandatory
 
  \abstract
   {Distance measurements to astronomical objects are essential for understanding their intrinsic properties. For asymptotic giant branch (AGB) stars it is particularly difficult to derive accurate distance estimates. Period-luminosity relationships rely on the correlation of different physical properties of the stars, while the angular sizes and variability of AGB stars make parallax measurements inherently inaccurate. For the carbon AGB star R~Sculptoris, the uncertain distance significantly affects the interpretation of observations regarding the evolution of the stellar mass loss during and after the most recent thermal pulse.}
   {We aim to provide a new, independent measurement of the distance to R~Sculptoris, reducing the absolute uncertainty of the distance estimate to this source.}
   {R~Scl is a semi-regular pulsating star, surrounded by a thin shell of dust and gas created during a thermal pulse $\approx$2000 years ago. The stellar light is scattered by the dust particles in the shell at a radius of $\approx$19\arcsec. The variation in the stellar light affects the amount of dust-scattered light with the same period and amplitude ratio, but with a phase lag that depends on the absolute size of the shell. We measured this phase lag by observing the star R~Scl and the dust-scattered stellar light from the shell at five epochs between June -- December 2017. By observing in polarised light, we imaged the shell in the plane of the sky, removing any uncertainty due to geometrical effects. The phase lag gives the absolute size of the shell, and together with the angular size of the shell directly gives the absolute distance to R~Sculptoris.}
   {We measured a phase lag between the stellar variations and the variation in the shell of 40.0$\pm$4.0 days. The angular size of the shell is measured to be 19\farcs1$\pm$0\farcs7. Combined, this gives an absolute distance to R~Sculptoris of 361$\pm$44 pc.}
   {We independently determined the absolute distance to R~Scl with an uncertainty of 12\%. The estimated distance is consistent with previous estimates, but is one of the most accurate distances to the source to date. In the future, using the variations in polarised, dust-scattered stellar light, may offer an independent possibility to measure reliable distances to AGB stars.}


   \maketitle
%
%-------------------------------------------------------------------

\section{Introduction}

The asymptotic giant branch (AGB) star R~Scl is a semiregular pulsating, carbon-rich star in the last stages of its life. During the AGB, low-mass stars shed most of their stellar envelopes in a strong, slow wind that enriches the interstellar medium with stellar processed material formed during thermal pulses (TPs). AGB stars contribute to the chemical evolution of galaxies and provide up to 70\% of the total dust budget~\citep{schneideretal2014}.

Estimating correct distances to AGB stars is notoriously difficult. Parallax measurements of nearby AGB stars are hampered by the fact that the size of the star itself, typically a few AU in radius, is comparable to or larger than the parallax. Additionally, bright features on the surface of the stars and/or varying extinction due to the dust circumstellar envelope lead to extremely unreliable parallaxes. Even the Gaia mission, which will measure parallaxes to stars in the Milky Way with unprecedented precision, will suffer from this problem. Phase lag measurements from OH masers, on the other hand, are limited to sources with OH maser emission, that is oxygen-rich stars~\citep[e.g.][]{vanlangeveldeetal1990}. \cite{groenewegenetal2012} determined the distance to the AGB star CW~Leo through phase lag measurements in features observed in thermal emission with Herschel/SPIRE. Although this estimate provides an independent measurement of the distance, it relies on an assumed geometry when deriving the absolute distance.

Studies of R~Scl show that it is possible to observationally study the evolution of the mass-loss rate and expansion velocity during and after a thermal pulse~\citep{maerckeretal2012,maerckeretal2016}. However, the estimated distance is a major source of uncertainty. It affects the measured absolute sizes of the detached shell, and hence all evolutionary timescales, for example the total age of the shell. It further plays an important role in the dust and molecular radiative transfer, affecting the total measured dust and molecular masses. 

We present observations with EFOSC2~\citep[ESO Faint Object and Spectrometer Camera, v.2;][]{buzzonietal1984} on the ESO/NTT telescope at La Silla observatory to image the detached shell of dust around the carbon AGB star R~Scl. The goal of the observations is to determine the distance to R~Scl by measuring the phase lag $\Delta t$ between the stellar variability and the variability in the detached shell due to dust-scattered stellar light. The shell around R~Scl was created during the brief increase in mass-loss rate and expansion velocity during a TP, and has a simple spherical geometry with a well-defined angular radius of $\approx$19\arcsec and a width of $\approx$2-3\arcsec~\citep{delgadoetal2001,delgadoetal2003a,olofssonetal2010,maerckeretal2014}. In polarised light, the shell appears as a ring~\citep{delgadoetal2003a,maerckeretal2014} with an absolute radius of $c\times\Delta t$, where $c$ is the speed of light and $\Delta t$ the measured phase lag~\citep[e.g.][]{sparks1994}. Combined with the angular size of the shell, this then directly gives an independent estimate of the distance to R~Scl.

In Sect.~\ref{s:observations} we present the EFOSC2 observations and describe the pulsation properties of R~Scl. Section~\ref{s:results} presents the measured variability in R~Scl and the detached shell, the resulting light curves, and the estimated distance. The results are discussed in Sect.~\ref{s:discussion}, and related to previous measurements of the distance to R~Scl. Conclusions and implications for future observations and other detached shell objects are presented in Sect.~\ref{s:conclusions}.

\section{Observations}
\label{s:observations}

\subsection{The light curve of R~Scl}
\label{s:lightcurve}

\begin{figure*}[t]
\centering
\includegraphics[width=12cm]{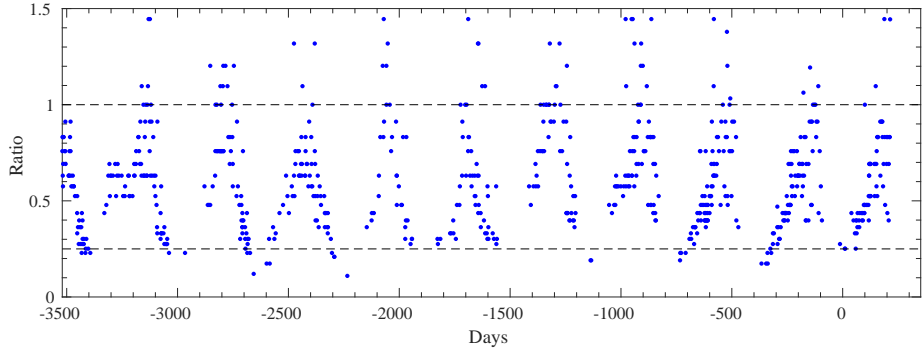}
\caption{Lightcurve of R~Scl based on V-band observations from AAVSO and relative to a magnitude of $m_V=$6.5 (i.e. ratio 1 in the figure). Observations are shown for the last ten years, relative to June 1st, 2016 (JD 2457540.5). The dashed lines indicate the minimum and maximum values corresponding to a min-to-max ratio of 0.25 (i.e. a magnitude amplitude of 0.75 magnitudes).}
\label{f:lightcurve}
\end{figure*}

\begin{figure}
\centering
\includegraphics[width=8cm]{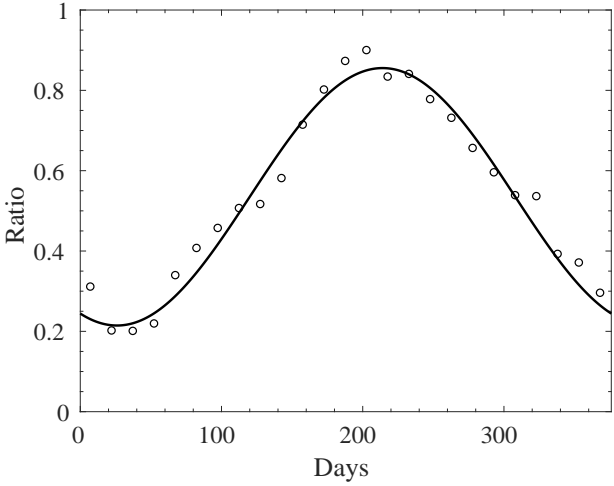}
\caption{Final ten periods folded and binned by 25 days (circles), together with a sinus function fit to the data. The sin-function has a period of 376 days and an amplitude of 0.75 magnitudes.}
\label{f:folded}
\end{figure}

R~Scl is a carbon-rich, semi-regular pulsating AGB star~\citep{samusetal2009}. \cite{wittkowskietal2017} fit a sine curve to the last ten pulsation periods, and derive a period of 376 days. Their curve has an amplitude of 0.7 mag in V-band. However, looking at the most recent periods (Fig.~\ref{f:lightcurve}), a magnitude amplitude in the pulsation of 0.75 mag appears more likely, corresponding to a min-to-max ratio in the brightness of 0.25. Using a sine curve to fit the pulsations reproduces the observed magnitudes reasonably well (Fig.~\ref{f:folded}). Variations in the light curve between different periods are difficult to predict, and the deviations from a sine curve lead to an added uncertainty in the estimated distance. We assume a sine curve with $P$=376 days and an amplitude of 0.75 mag for fitting the observed light curve in R~Scl and the shell.

\subsection{EFOSC observations}
\label{s:efoscobs}

 In order to measure the variability in the star, R~Scl was observed during five epochs between June 2016 and December 2016, that is sampling approximately half a pulsation period of R~Scl (see Sect.~\ref{s:lightcurve}). For each epoch, observations were scheduled to be carried out on three consecutive nights to ensure that useable data was obtained for each epoch. Given the sky conditions, we managed to observe useful data during one night in Epoch 1, and two nights in Epochs 2 and 4. Epochs 3 and 5 have the full three nights of measurements. EFOSC2 was used in direct imaging, polarisation, and coronagraphic mode, using the standard ESO V-band filter (\veso), a customised neutral density filter, and a customised V-band filter including an adhesive polarising film (Table~\ref{t:filters}). The pixel scale of EFOSC2 on the NTT is 0\farcs24$\times$0\farcs24. The field of view is 4\arcmin$\times$4\arcmin. 

\begin{figure}
\centering
\includegraphics[width=8cm]{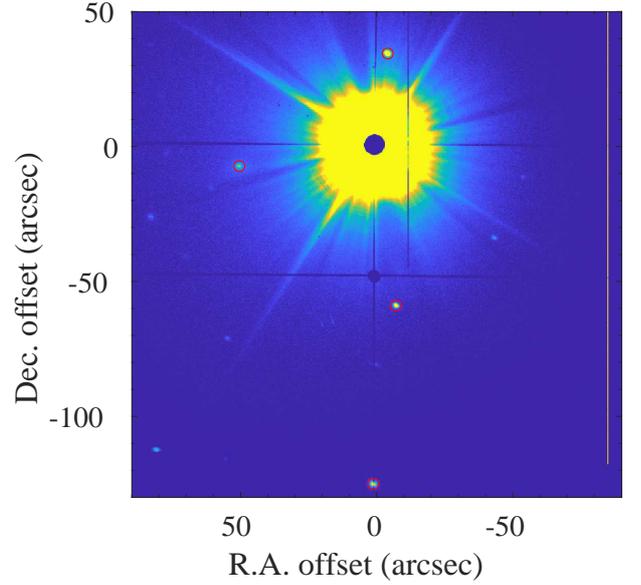}
\caption{180\arcsec$\times$180\arcsec~part of the field observed with EFOSC2 in the \veso~ filter. The background stars used to scale the brightness of R~Scl and the shell are marked with red circles.}
\label{f:bgfield}
\end{figure}

In order to eliminate the uncertainty due to absolute flux calibration, we measured the variation of the star and the detached shell relative to four background stars spread throughout the field of view (Fig.~\ref{f:bgfield}). This reduces any uncertainty due to different pointings on the sky for observations of flux calibrators, decreasing the uncertainty in our distance estimate.

\begin{table}
\caption{Filters and coronagraphs used in the observations, central wavelengths ($\lambda$) and filter widths ($\Delta \lambda$)}
\begin{center}
\begin{tabular}{l c c l}
\hline\hline
Filter & $\lambda$ & $\Delta \lambda$ & Comments\\
      & [nm] & [nm] &\\
\hline
V-pol & 550 & 88 & including adhesive\\
         &         &        & polarising film\\
\veso & 548 & 113 & --\\
ND3 & -- & -- & neutral density filter, \\
         &    &     &reduction by 1000\\
coronagraph & -- & -- & 8\arcsec coronagraphic spot\\
\hline\hline
\end{tabular}
\end{center}
\label{t:filters}
\end{table}%

\subsection{Direct observations of R~Scl}
\label{s:rscldirect}

We observed R~Scl directly using the ESO V-band filter (\veso), centred at $\lambda_V=547.6$\,nm with a width $\Delta\lambda_V=113.2$\,nm. In V-band, R~Scl has a peak magnitude of $m_V\approx6.5$ (Fig.~\ref{f:lightcurve}). In direct observations through broadband filters, EFOSC2 will saturate immediately for a star with this magnitude. In order to reduce the direct stellar light, we observed R~Scl through a neutral density filter (ND3), reducing the brightness of the star by a factor of $10^3$. The ND3 filter was entered in the grism wheel, in front of the \veso~filter. Observations combining the ND3 and \veso~filters were used to measure the variation in R~Scl directly.

\subsection{Polarised observations of the detached shell}
\label{s:shellobs}

In order to accurately measure the phase lag to R~Scl, it is necessary to know the correct angular size of the detached shell. This is best done in polarised dust-scattered stellar light. The stellar light is polarised as it is scattered by the dust grains in the detached shell, and the polarisation is strongest when the angle between the incident light and scattered light is 90 degrees. In polarised light the detached shell hence appears as a ring, with a radius that corresponds to the angular size of the shell. This method has been used repeatedly to observe the detached shells around carbon stars~\citep{delgadoetal2003a,maerckeretal2010,maerckeretal2014,ramstedtetal2011}. The intensity of the polarised light varies with the brightness of the central star, with the same period and the same ratio between minimum and maximum brightness. By using polarised light observations, we therefore can both accurately measure the angular size of the shell, as well as the variation in intensity in the shell. 

Without the ND3 filter, the brightness of the central star requires the use of a coronagraph. EFOSC2 offers two coronagraphic spots with diameters of 4\arcsec\,and 8\arcsec. The shell around R~Scl has a radius of $\approx$20\arcsec, and we used the larger of the two spots to effectively block the direct stellar light without affecting the measurements of the shell (Fig.~\ref{f:rsclIall}).

The usual setup of EFOSC2 in polarimetric mode uses an imaging filter, a half-wave or quarter-wave plate, and a wollaston prism splitting the image into two strips 10\arcsec\,or 20\arcsec\,wide, with one polarisation each. All these elements together take up the available filter and grism wheels in EFOSC2, and it is not possible to use the coronagraphic spots in polarisation mode in any standard setup. Additionally, since the shell around R~Scl has a radius of $\approx$20\arcsec, the splitting of the image into narrow strips is not ideal, making it impossible to cover the entire shell in one image.

In order to avoid the problems with the standard polarisation setup, we used a customised V-band filter, with an adhesive, polarising film (\vpol). This made it possible to use EFOSC2 in standard coronagraphic mode using the half-wave plate, and with the \vpol~filter inserted in the filter wheel. The observations provided polarised images at four polarisation angles (0\textdegree, 45\textdegree, 90\textdegree, and 135\textdegree). These could then be combined to produce images in total and polarised intensity~\citep[e.g.][]{maerckeretal2010,maerckeretal2014,ramstedtetal2011}. The polarised intensity images were used to measure the variation of the scattered light in the shell (Fig.~\ref{f:rsclPall}).

\subsection{Observing strategy}
\label{s:strategy}

The variation in brightness of the star and shell is measured relative to background stars in the image. Ideally, the background stars are measured in the same image as the star and shell, reducing any uncertainties due to changing sky conditions. However, due to the strong contrast in brightness between R~Scl, the background stars, and the shell, this was not possible. The use of the ND3 filter for the direct observations of R~Scl makes it impossible to also detect the shell or background stars in the same image. In order to reduce the uncertainty due to varying sky conditions, the background stars were observed between observations in polarisation mode and observations with the ND3 filter. A typical sequence of observations is presented in Table~\ref{t:obsset}. 

\begin{table*}
\caption{Typical observing sequence for R~Scl. Exposure times are representative values typically used, but were adjusted based on the prevailing observing conditions. Per observing night two to three sequences were observed.}
\begin{center}
\begin{tabular}{c l c c c l}
\hline\hline
Set & Filter & Coronagraph & Polarisation & Exposures & Product\\
\hline
1 & V-pol & yes & yes & 6$\times$30s & Polarised intensity from shell\\
2 & \veso & yes & no & 6$\times$30s & Total intensity from background stars\\
3 & \veso + ND3 & no & no & 3$\times$15s & Total intensity from R~Scl\\
\hline\hline
\end{tabular}
\end{center}
\label{t:obsset}
\end{table*}%

Domeflats were observed during the day by the daytime operator. All images were then flatfielded, background subtracted, clipped, median combined, and scaled to 1s to give coronagraphic images at each polarisation angle in the \vpol~filter, coronagraphic images in total intensity in the \veso filter, and direct images of R~Scl in the \veso + ND3 filters in counts/s.

\begin{figure}
\centering
\includegraphics[width=6cm]{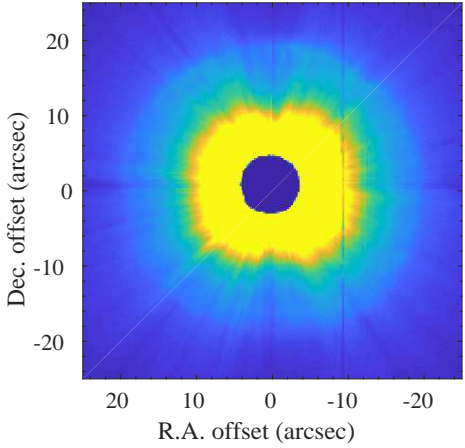}
\caption{All epochs of R~Scl combined in total intensity.}
\label{f:rsclIall}
\end{figure}

\begin{figure}
\centering
\includegraphics[width=6cm]{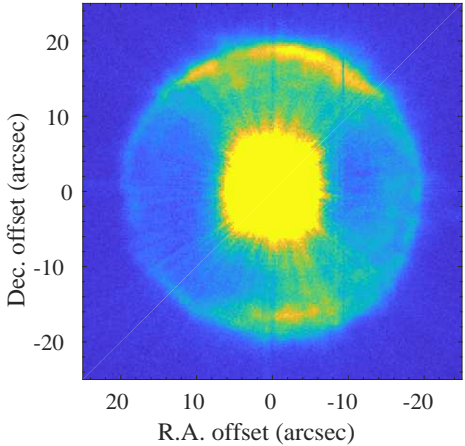}
\caption{All epochs of R~Scl combined in polarised intensity.}
\label{f:rsclPall}
\end{figure}

Depending on the visibility of R~Scl throughout the observing period, we usually managed to observed 2-3 sets per night. In some epochs not all sets could be used, due to varying weather conditions. Figures~\ref{f:rsclIall} and~\ref{f:rsclPall} show all epochs of R~Scl combined in total intensity and polarised intensity, respectively.

Flux standards were also observed for a basic consistency check, confirming the magnitude of R~Scl. Since the absolute calibration is less certain than the relative calibration against background stars, and not required in this study, we do not present any absolute flux calibration.

\section{Results}
\label{s:results}

\subsection{Brightness ratios}
\label{s:brightness}

    The brightness of the point sources (i.e. R~Scl and the background stars) was measured through aperture photometry in the ND3+\veso~and \veso~images. The apertures were 20 pixels and 15 pixels for R~Scl and the background stars, respectively. The sky background was subtracted, defined by the average of counts in annuli of 25-35 pixels and 16-25 pixels for R~Scl and the background stars, respectively. The resulting brightness is given in counts/s for R~Scl ($I_{\rm{RScl}}$) and the background stars ($I_{\rm{bg}}$), and the uncertainty for each measurement ($\sigma_{\rm{RScl}}$ and $\sigma_{\rm{bg}}$) is determined by the rms in the background region. An average value was calculated for the four background stars. The results are presented in Table~\ref{t:obsmeasurements}.

\begin{figure}
\centering
\includegraphics[width=6cm]{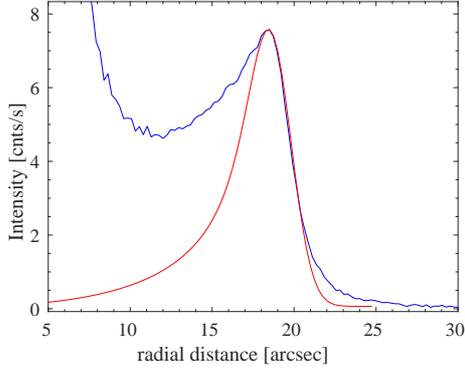}
\caption{Azimuthally averaged radial profile of the observed polarised intensity from the second set of observations on December 15, 2016 (blue). The expected radial profile from a shell with $R$=19.2\arcsec and FWHM=2\farcs7 (red).}
\label{f:polprof}
\end{figure}

For the shell we created azimuthally averaged radial profiles of the polarised intensity for each observation set. The radial profiles have a spatial resolution of one pixel (i.e. 0\farcs24). Figure~\ref{f:polprof} shows an example of such a profile from December 2016. The peak at 18\farcs5 due to the detached shell can clearly be seen, and is consistent with previous, similar observations with PolCor~\citep{maerckeretal2014}. The increase in emission at smaller radii is due to the remaining stellar psf, more recent mass loss, and additional scattering at the edge of the mask. Due to limb brightening in the shell, the true peak of the shell lies at a slightly larger radius than the peak in the radial profile. Figure~\ref{f:polprof} shows the theoretical profile expected from a thin shell assuming a Gaussian density distribution with $R$=19\farcs2 and FWHM=2\farcs7 (the profile only considers the shell, i.e. no emission from inside the shell due to more recent mass loss). The theoretical profile is calculated using an analytical formula for dust scattering, assuming isotropic scattering by the dust grains~\citep{henyeyco1941,draine2003,maerckeretal2010}.

The lightcurve of the shell is measured by fitting a theoretical profile to the observed radial profiles, adjusting the radius and peak intensity $I_{\rm{shell}}$ to determine two parameters: the peak intensity in the radial profile and the radius of the Gaussian distribution. This method has been used previously for R~Scl and other detached shell sources~\citep{maerckeretal2010,maerckeretal2014}. We note that the shell is not perfectly symmetric. In particular, there is a flattening of the shell in the south. This has been observed in previous observations in dust scattered light, and is consistent with observations in CO line emission~\citep{maerckeretal2012,maerckeretal2014,maerckeretal2016}. The asymmetric structures mainly affect the measured width of the shell, and not the average radius. Excluding the flattened feature from the average profile increases the measured radius by $\approx$0\farcs1, significantly less than the assumed error in the measured average radius~\citep[Sect.~\ref{s:uncertainties}, and][]{maerckeretal2014}. The error in the peak brightness $\sigma_{\rm{shell}}$ is determined by the rms value in that radial bin (defined as $\sigma_{\rm{Im}}/\sqrt{N_{\rm{bin}}}$, where $N_{\rm{bin}}$ is the number of bins included in the average value of that radial bin and $\sigma_{\rm{Im}}$ is the overall rms across the image). The measured peak intensities with uncertainties and radii are presented in Table~\ref{t:obsmeasurements}.

For each set of observations the ratio between measurements was formed

\begin{equation}
\label{e:ratiostar}
r_{\rm{RScl}}={I_{\rm{RScl}} \over I_{\rm{bg}}},
\end{equation}

and the uncertainty in the ratio was determined through standard error propagation:

\begin{equation}
\label{e:eratiostar}
\sigma_r=r_{\rm{RScl}}\times \sqrt{{\sigma^2_{\rm{RScl}} \over I_{\rm{RScl}}^2} + {\sigma^2_{\rm{bg}} \over I_{\rm{bg}}^2} }.
\end{equation}

And the equivalent was done for $I_{\rm{shell}}$. The ratios of each epoch are averaged to give the final ratios $R_{\rm{RScl}}$ and $R_{\rm{shell}}$ (with respective errors) for the star and the shell, respectively.

\begin{table*}[htp]
\caption{Observations of R~Scl, four background stars, the polarised shell, and the shell radius. The measurements are given in counts/s. The days are relative to June 1, 2016 (JD 2457540.5). }
\begin{center}
\tiny
\begin{tabular}{l l c c c c c c c c c c c c c c c}
\hline\hline
Day & Set & \multicolumn{2}{c}{R~Scl}& \multicolumn{2}{c}{BG-S1} &\multicolumn{2}{c}{BG-S2} &\multicolumn{2}{c}{BG-E} &\multicolumn{2}{c}{BG-N}&\multicolumn{2}{c}{BG-AVG} & \multicolumn{2}{c}{Shell}& Radius\\
 & & $I_{\rm{RScl}}$ & $\sigma_{\rm{RScl}}$&$I_{bg-S1}$ & $\sigma_{bg-S1}$&$I_{bg-S2}$ & $\sigma_{bg-S2}$&$I_{bg-E}$ & $\sigma_{bg-E}$&$I_{bg-N}$ & $\sigma_{bg-N}$&$I_{\rm{bg}}$ & $\sigma_{\rm{bg}}$&$I_{\rm{shell}}$ & $\sigma_{\rm{shell}}$& [\arcsec]\\
\hline
\multicolumn{4}{l}{\textbf{Epoch 1 -- June}}&&&&&&&&&&&&&\\
16	&	set 2	&	41622	&	9.0	    &	1911.9	&	602.3	&	1338.4	&	611.9	&	1002.2	&	637.7	&	3777.3	&	643.6	&	2007	&	312	&	1.5	&	0.04	&	19.0\\
\multicolumn{2}{l}{\it{Ratio:}}&\multicolumn{3}{l}{$R_{\rm{RScl}}=20.7\pm3.2$}&\multicolumn{3}{l}{$R_{\rm{shell}}=0.75\pm0.11$}&&&&&&&&&\\
\hline
\multicolumn{4}{l}{\textbf{Epoch 2 -- July}}&&&&&&&&&&&&&\\
46	&	set 1	&	50555	&	13.2	&	1593.8	&	43.6	&	1269.2	&	24.3	&	829.8	&	39.4	&	2892.3	&	97.2	&	1646	&	29	&	1.8	&	0.06	&	19.0\\
	&	set 2	&	60003	&	14.0	&	1960.5	&	27.4	&	1452.4	&	14.9	&	978.9	&	31.0	&	4008.3	&	92.1	&	2100	&	26	&	1.5	&	0.05	&	19.5\\
48	&	set 1	&	58118	&	15.5	&	1375.9	&	63.6	&	1254.5	&	36.4	&	872.0	&	44.9	&	3664.4	&	190.5	&	1792	&	52	&	1.5	&	0.10	&	19.0\\
	&	set 2	&	59887	&	14.8	&	1705.9	&	45.2	&	1378.8	&	21.9	&	912.3	&	42.3	&	3918.7	&	145.6	&	1979	&	40	&	1.6	&	0.07	&	19.0\\
	&	set 3	&	61277	&	12.9	&	1724.5	&	39.5	&	1434.5	&	19.9	&	911.4	&	43.6	&	4365.8	&	172.3	&	2109	&	46	&	1.9	&	0.08	&	19.0\\
\multicolumn{2}{l}{\it{Ratio:}}&\multicolumn{3}{l}{$R_{\rm{RScl}}=30.1\pm0.3$}&\multicolumn{3}{l}{$R_{\rm{shell}}=0.88\pm0.02$}&&&&&&&&&\\
\hline
\multicolumn{4}{l}{\textbf{Epoch 3 -- September I}}&&&&&&&&&&&&&\\
96	&	set 1	&	95159	&	23.7	&	2232.4	&	27.2	&	1540.5	&	11.8	&	903.9	&	54.4	&	5374.8	&	132.7	&	2513	&	37	&	4.4	&	0.10	&	19.1\\
	&	set 2	&	95814	&	27.7	&	2098.7	&	27.6	&	1444.7	&	12.1	&	893.8	&	37.8	&	4345.2	&	136.0	&	2196	&	36	&	4.4	&	0.10	&	19.1\\
97	&	set 1	&	99705	&	23.5	&	2161.0	&	33.8	&	1567.8	&	18.0	&	965.5	&	41.6	&	4633.3	&	143.6	&	2332	&	39	&	4.4	&	0.09	&	19.2\\
	&	set 2	&	98577	&	22.2	&	2184.2	&	28.2	&	1369.9	&	13.7	&	946.6	&	45.4	&	3345.6	&	121.3	&	1962	&	33	&	4.5	&	0.11	&	19.2\\
98	&	set 1	&	98347	&	34.6	&	2158.0	&	35.1	&	1546.2	&	17.6	&	944.4	&	37.3	&	4511.4	&	128.5	&	2290	&	35	&	4.5	&	0.10	&	19.2\\
	&	set 2	&	98382	&	32.0	&	2206.6	&	27.5	&	1385.2	&	12.3	&	1047.7	&	41.2	&	4011.0	&	115.4	&	2163	&	32	&	4.7	&	0.11	&	19.1\\
\multicolumn{2}{l}{\it{Ratio:}}&\multicolumn{3}{l}{$R_{\rm{RScl}}=43.8\pm0.3$}&\multicolumn{3}{l}{$R_{\rm{shell}}=2.01\pm0.02$}&&&&&&&&&\\
\hline
\multicolumn{4}{l}{\textbf{Epoch 4 -- September II}}&&&&&&&&&&&&&\\
107	&	set 1	&	103100	&	31.7	&	1535.9	&	104.5	&	1344.4	&	48.0	&	911.2	&	82.5	&	3524.7	&	250.7	&	1829	&	72	&	4.4	&	0.12	&	19.1\\
	&	set 2	&	102010	&	31.7	&	1722.3	&	79.0	&	1448.7	&	38.3	&	813.2	&	80.0	&	3305.5	&	252.2	&	1822	&	70	&	4.6	&	0.13	&	19.0\\
	&	set 3	&	100470	&	28.3	&	1562.2	&	85.0	&	1448.9	&	29.5	&	1031.1	&	60.3	&	4349.7	&	287.5	&	2098	&	77	&	4.6	&	0.12	&	19.0\\
108	&	set 1	&	101510	&	36.0	&	1901.2	&	79.0	&	1509.1	&	37.7	&	1053.3	&	71.9	&	3653.6	&	201.7	&	2029	&	58	&	4.7	&	0.13	&	19.0\\
	&	set 2	&	103140	&	41.3	&	1943.3	&	75.1	&	1565.3	&	33.8	&	891.8	&	67.7	&	4892.4	&	219.1	&	2323	&	61	&	4.9	&	0.12	&	19.1\\
	&	set 3	&	102920	&	47.9	&	1990.6	&	64.7	&	1491.1	&	29.9	&	908.8	&	70.4	&	3209.8	&	196.1	&	1900	&	55	&	5.0	&	0.13	&	19.1\\
	&	set 4	&	98752	&	54.2	&	1855.4	&	74.2	&	1518.3	&	28.4	&	1111.9	&	55.4	&	4435.4	&	246.7	&	2230	&	66	&	4.7	&	0.13	&	19.1\\
\multicolumn{2}{l}{\it{Ratio:}}&\multicolumn{3}{l}{$R_{\rm{RScl}}=50.8\pm0.7$}&\multicolumn{3}{l}{$R_{\rm{shell}}=2.33\pm0.04$}&&&&&&&&&\\
\hline
\multicolumn{4}{l}{\textbf{Epoch 5 - December}}&&&&&&&&&&&&&\\
195	&	set 1	&	125730	&	43.8	&	1680.1	&	65.8	&	1235.0	&	29.9	&	820.5	&	68.3	&	2892.6	&	160.1	&	1657	&	47	&	7.7	&	0.16	&	19.1\\
	&	set 2	&	117520	&	46.3	&	1483.4	&	85.7	&	1290.7	&	28.4	&	1028.3	&	52.3	&	3935.4	&	203.7	&	1934	&	57	&	7.5	&	0.16	&	19.2\\
196	&	set 1	&	119830	&	29.9	&	1807.3	&	52.6	&	1241.1	&	18.9	&	911.9	&	50.1	&	3778.4	&	163.3	&	1935	&	45	&	7.6	&	0.15	&	19.2\\
	&	set 2	&	113920	&	28.5	&	1663.1	&	69.5	&	1282.3	&	22.3	&	805.3	&	50.4	&	4101.3	&	187.8	&	1963	&	52	&	7.4	&	0.15	&	19.2\\
197	&	set 1	&	122550	&	42.3	&	2128.5	&	43.5	&	1258.3	&	18.1	&	745.0	&	57.2	&	2517.8	&	150.2	&	1662	&	42	&	8.2	&	0.16	&	19.2\\
	&	set 2	&	116960	&	40.9	&	1616.1	&	52.6	&	1328.2	&	20.6	&	738.9	&	49.5	&	4239.7	&	180.8	&	1981	&	49	&	7.6	&	0.15	&	19.2\\
\multicolumn{2}{l}{\it{Ratio:}}&\multicolumn{3}{l}{$R_{\rm{RScl}}=64.9\pm0.7$}&\multicolumn{3}{l}{$R_{\rm{shell}}=4.16\pm0.06$}&&&&&&&&&\\

\hline\hline
\end{tabular}
\end{center}
\label{t:obsmeasurements}
\end{table*}%

\begin{figure}
\centering
\includegraphics[width=8cm]{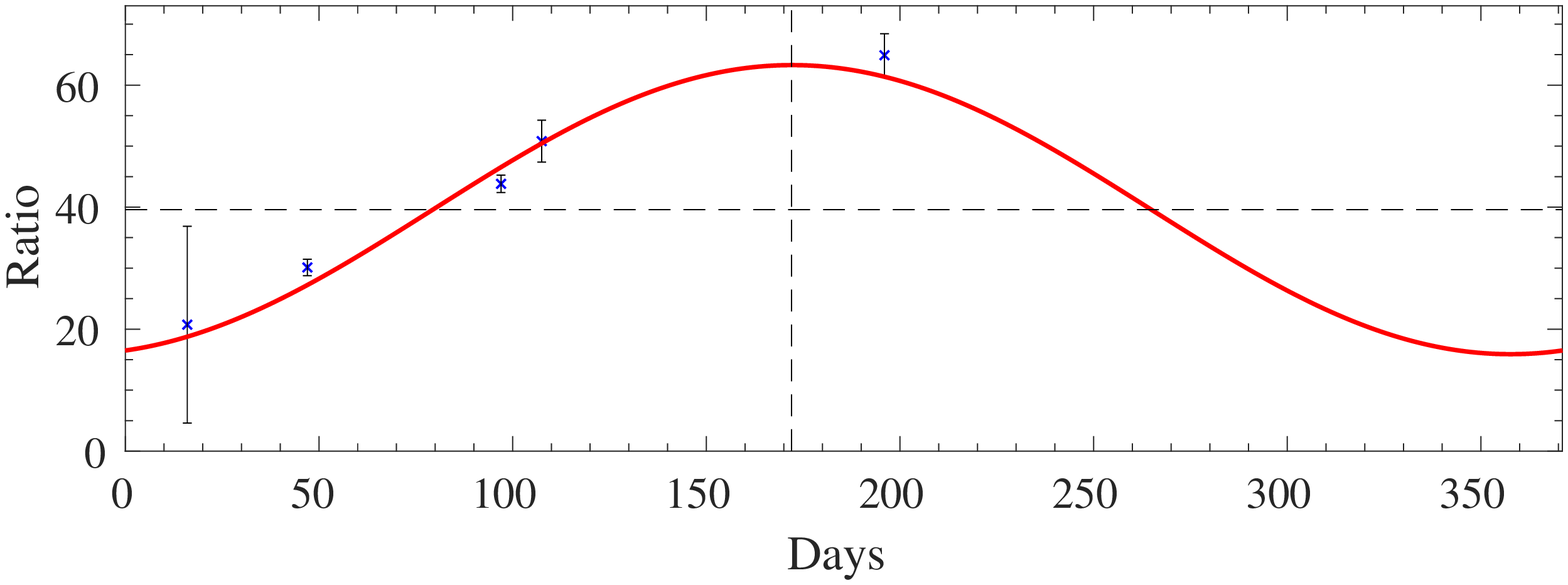}
\includegraphics[width=8cm]{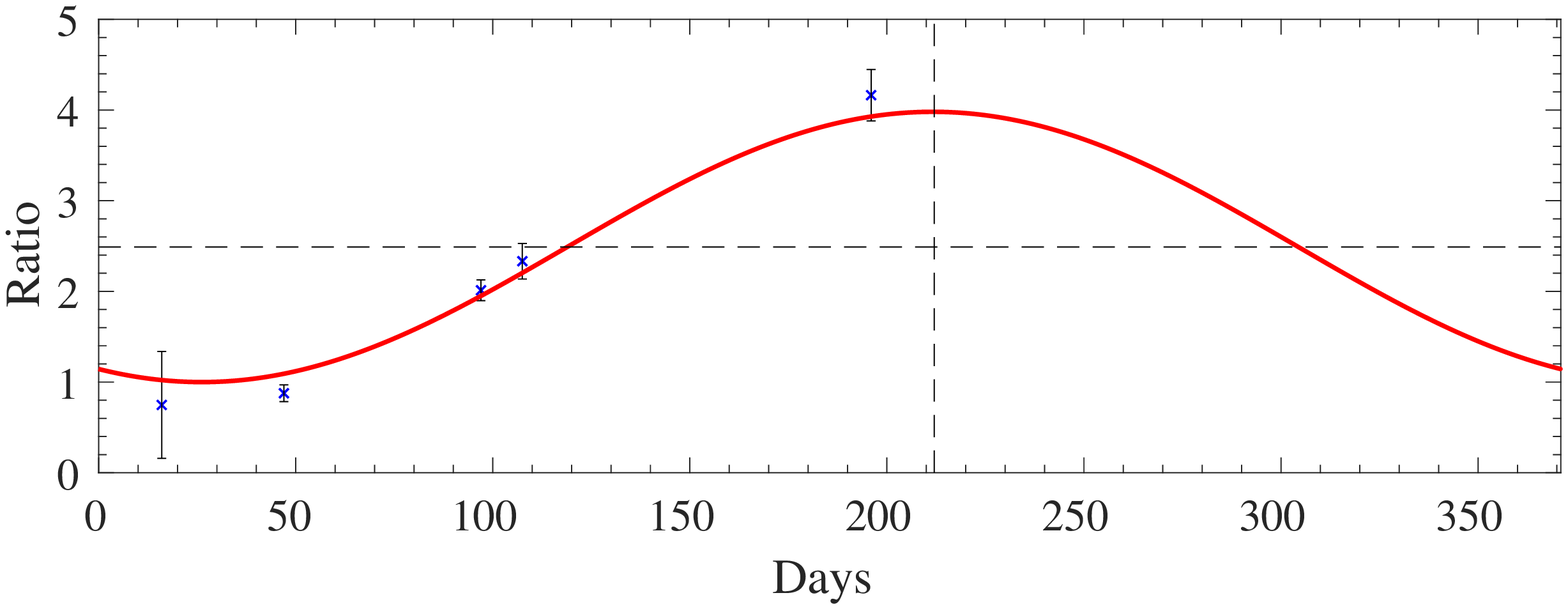}
\caption{Observed light curves for R~Scl (top) and the detached shell (bottom). The days are relative to June 1, 2016 (JD 2457540.5). The intensity of R~Scl is reduced by a factor 1000 by the ND3 filter. The values of the ratio for the shell however were multiplied by a factor 1000 for the sake of readability. For both panels the errors were multiplied by five for visualisation. The dashed lines show the best-fit ratio and peak of the light curve for this period.}
\label{f:obscurves}
\end{figure}

\subsection{Lightcurve fitting}
\label{s:curvefit}

In order to determine the phase lag between the stellar pulsations and the brightness of the shell, we fit a sine curve to the observed data points. The sine curve is fixed to have a period of 376 days, and minimum to maximum ratio of 0.25. This leaves two free parameters when fitting the sine curve to the observations: a shift in days (i.e. the x-direction), and a shift in the average ratio (i.e. the y-direction). The difference in shift in days between the light curves fit to $R_{\rm{RScl}}$ and $R_{\rm{shell}}$ gives the phase lag, and hence the absolute size of the shell. Together with the angular size of the shell, this gives the distance to R~Scl.

Figure~\ref{f:obscurves} shows the observed light curves for R~Scl and the shell. The sources of uncertainties in the measured ratios are discussed in more detail in Sect.~\ref{s:uncertainties}.

\begin{figure}[t]
\centering
\includegraphics[width=6cm]{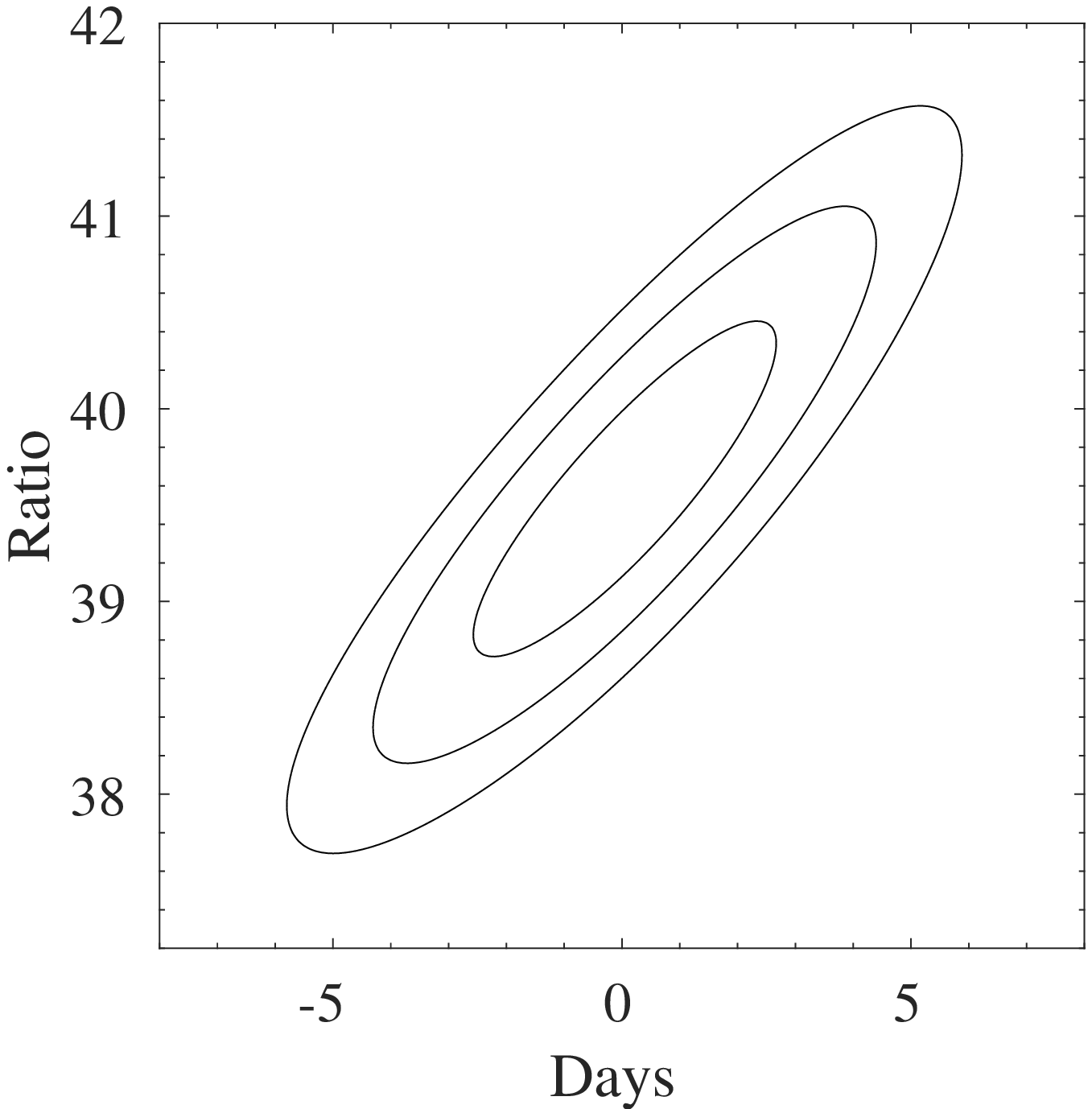}
\includegraphics[width=6cm]{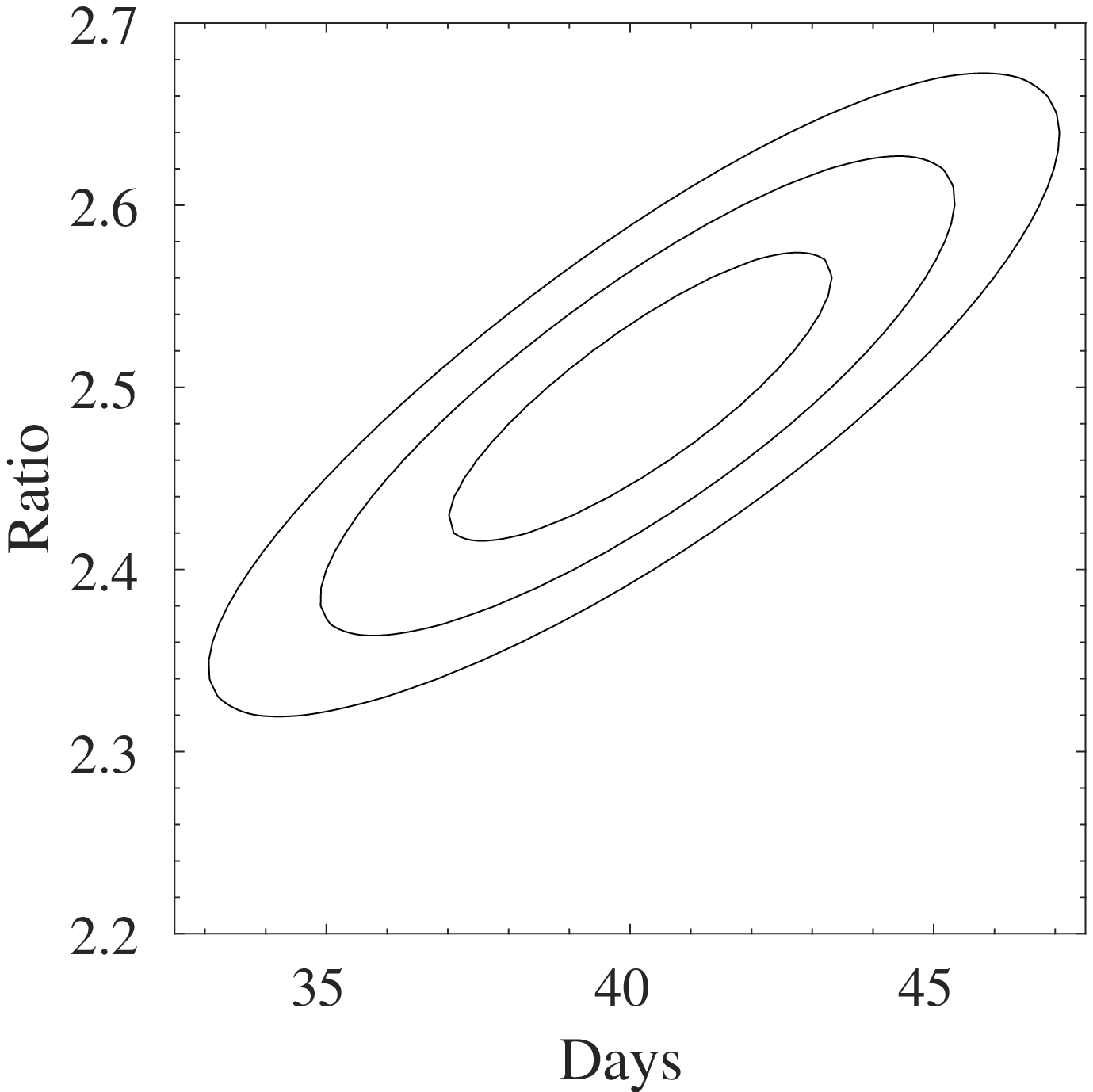}
\caption{$\chi^2$ maps from sine curves fit to the observed ratios for R~Scl (top) and the shell (bottom). The shift in days is given relative to the best-fit model for R~Scl.}
\label{f:chi2maps}
\end{figure}

In order to fit the sine curve to the data points, we created grids of sine curves varying the days and average ratios, and determine the $\chi^2$-value for each fit:

\begin{equation}
\label{e:chi2}
\chi^2={1\over N_i} \times \sum {(R_i-F_i)^2 \over\sigma_{R,i}},
\end{equation}

where $N_i$ is the number of points, $R_i$ are the observed ratios, $F_i$ the model value in point $i$, and $\sigma_{r,i}$ the uncertainty in the observed ratio. The resulting $\chi^2$-maps are shown in Fig.~\ref{f:chi2maps}. The uncertainty in the shift in days for the light curve of R~Scl and the shell are $\pm2.6$ and $\pm3.1$ days, respectively. The resulting shift in the light curves between R~Scl and the shell is 40.0$\pm$4.0\,days, corresponding to an absolute distance between the star and shell of 6900$\pm$690\,AU. The average angular size of the measured peak is 19\farcs1$\pm$0\farcs1. However, comparing to earlier measurements of the shell radius~\citep[e.g.][and references therein]{maerckeretal2014}, the uncertainty in the shell radius is likely larger. Here we adopt the uncertainty of $\pm$0\farcs7 from~\cite{maerckeretal2014}. This leads to a distance to R~Scl of 361\,pc, with a measurement uncertainty of $\pm$39\,pc. The absolute uncertainty is likely higher (Sect.~\ref{s:uncertainties}).

\section{Discussion}
\label{s:discussion}

\subsection{Observational uncertainties}
\label{s:uncertainties}

The uncertainty in the measured fluxes for R~Scl in the ND3+\veso~filter combination is generally $\approx0.05\%$. The uncertainty in the measured flux averaged between the four background stars in the \veso~filter is $\approx$2\%. With the exception of the first epoch (June 17, 2016) the uncertainty in the ratio $R_{\rm{RScl}}$ is $\approx1\%$ for each epoch. The one night of observations in Epoch~1 still had unstable weather, leaving us with only one set of useful observations, and an uncertainty of $\approx$15\% for this measurement. The uncertainty in the measured peak intensities in polarised light of the shell is $\approx2-4\%$. The error in the measured ratios $R_{\rm{shell}}$ is  $\approx2\%$.

In principle it would have been possible to measure the brightness of the background stars directly in the total intensity images of the \vpol~observations (making them simultaneous with the polarised observations). However, the transmission of the \vpol~filter with the polarising film applied was $\approx$50\% of the \veso\, filter, reducing the signal-to-noise of these observations and hence increasing the relative error to $\sim$10\%. We therefore chose to use the \veso~observations of the background stars also for the ratios with the shell emission. 

The total error in the distance due to these observational uncertainties lies at 11\%. Additional uncertainties that can affect the distance estimate are the assumed pulsation period and amplitude. Decreasing and increasing the pulsation period by 5 days changes the estimated distance to 357 pc and 368 pc, respectively. A change in amplitude in the fitted sine curves to 0.7 mag and 0.8 mag gives best-fit distances of 340 pc and 380 pc, respectively. The true uncertainty in the distance is therefore likely larger than 39~pc. Using the range of possible distances due to changes in the period and amplitude as an indication, we add uncertainties of $\pm$5 pc and $\pm$20 pc, respectively. Adding these uncertainties in quadrature, the total uncertainty in the estimated distance then becomes 44 pc. 

The fractional uncertainty in the distance to R~Scl (12\%) is comparable to the uncertainty in the distance to CW~Leo, also determined through phase lag measurements~\citep[123$\pm$14 pc, that is 11\%;][]{groenewegenetal2012}, with the difference that our method does not have to make any assumptions on the geometry of the source. Adding more epochs to the observations of R~Scl may somewhat reduce the fractional uncertainty. Adding two more observations approximately one year after our observations, and assuming similar observational uncertainties, reduces the fractional uncertainty to $\approx8\%$. In particular, the effect of the uncertainty in the amplitude of the variation is reduced to $\pm$10~pc.

An unfortunate fact is that we had to use a neutral density filter to observe R~Scl directly, and that the total intensity observations through the \vpol~filter had a relatively large error compared to the \veso~observations. Ideally one would measure R~Scl directly, the polarised shell, and the background stars in the same exposure. 

When observing the polarised, dust-scattered stellar light around AGB stars with the PolCor instrument~\citep{ramstedtetal2011,maerckeretal2014}, the coronagraphic spots were neutral density spots that would reduce the direct stellar light to different degrees. In PolCor observations this allows for a careful placement of the star behind the mask, reducing the artefacts in the images. In these observations, one cycle of all polarisations angles would allow us to create images in polarised and total intensity that would contain the direct star light, the shell, and the background stars from the same exposures. Unfortunately the field of view of PolCor is too small to contain any of the background stars used here. 

\subsection{Comparison to other distance estimates to R~Scl}
\label{s:compare}

Table~\ref{t:prevdist} presents earlier estimates of the distance to R~Scl, mainly based on ($P$-$L$) relationships. All measurements, with the exception of the Hipparcos distance, lie within our estimated distance of 361 pc $\pm$ 44 pc, including the estimate based on SED fitting to hydrostatic atmosphere models by~\cite{sacutoetal2011}. Interestingly, the Hipparcos distance is the only one that is not consistent with any of the other measurements, but has the smallest uncertainty. The difficulty with measuring distances to AGB stars using parallaxes is that the angular size of the star is comparable to the parallax itself. For a diameter of 2 AU, the stellar disk of R~Scl would have an angular size of $\approx$5 mas at the distance of 380 pc, while the revised parallax is only 3.76 mas. Time-variable features on the stellar disk, and/or obscuration due to the dusty circumstellar envelope, might easily disturb the measured parallax. Most recently, the stellar disk of R~Scl has been observed and shown to exhibit such asymmetric features~\citep{wittkowskietal2017}.

In general, ($P$-$L$) relationships give relatively large errors, and rely on empirical relationships for large samples of sources. This increases the uncertainty for individual cases. Fitting the SED to hydrostatic models of the atmosphere treats R~Scl individually. However, this method relies on assumptions about the stellar atmosphere, and the underlying model uncertainties.

Our distance estimate is completely independent from the methods used by previous estimates. Further, the method itself does not make very critical or complicated assumptions, and is very straight-forward to measure and apply. The uncertainty of 12\% is one of the lowest uncertainties measured, along with the distance determined through hydrostatic atmosphere models~\citep{sacutoetal2011}. 

\begin{table*}[htp]
\caption{Distance estimates to R~Scl from the literature compared to our estimate.}
\begin{center}
\begin{tabular}{c c l  l}
\hline\hline
$D$ & $\sigma D$ & Method & Reference\\ 
 $[\rm{pc}]$ & [pc] & & \\
\hline
360 & -- & ($P$-$L$) relationship for Mira variables & \cite{groenewegenco1996}\\
370 & -70 / +100 & ($P$-$L$) relationship for SR variables & \cite{knappetal2003}\\
340 & -70 / +100 & ($P$-$L$) relationship for Mira variables & \cite{knappetal2003}\\
266 & -45 / +66 & revised Hipparcos & \cite{vanleeuwen2007}\\
370 & -- & ($P$-$L$) relationship for Mira variables & \cite{whitelocketal2008}\\
350 & -50 / +50 & SED fitting to hydrostatic atmosphere models & \cite{sacutoetal2011}\\
& & & \\
\textbf{361} & \textbf{-44 / + 44} & \textbf{Phase lag measurements in the detached shell }& \textbf{This paper}\\
\hline\hline
\end{tabular}
\end{center}
\label{t:prevdist}
\end{table*}%

\section{Conclusions}
\label{s:conclusions}

We have provided an independent distance estimate to the carbon AGB star R~Scl. At 361 pc $\pm$ 44 pc, our distance is consistent with most previous distance estimates, however with a lower uncertainty. The absolute size of the detached shell around R~Scl was measured by determining the phase lag between brightness variations in the star due to the stellar pulsations to the variation in brightness in the detached shell due to dust-scattered stellar light. By observing the shell in polarised light, geometrical uncertainties are eliminated, as the scattering will polarise the light preferentially in the plane of the sky, causing the detached shell to appear as a well-defined ring in the polarised images. By measuring the variation in intensity relative to background stars in the same field of view, calibration uncertainties are also removed. Except for the pulsation period and amplitude, the method does not make any assumptions regarding the characteristics of the star. As such, it provides one of the most reliable distance estimates to date. The agreement between our estimated distance, and the distance estimate through SED fitting of hydrostatic atmosphere models~\citep{sacutoetal2011}, adds to the reliability of both methods.

With this distance in hand, one of the largest uncertainties in models of the shell around R~Scl is reduced. The assumed distance to R~Scl affects all absolute sizes and time scales derived from the observations of the detached shell~\citep{maerckeretal2012,maerckeretal2016}. Only with a correct distance is it possible to properly constrain hydrodynamical models of the evolution of the shell, and hence its connection to thermal pulses and the evolution of the wind throughout the thermal-pulse cycle.

We are currently developing a new version of the PolCor instrument with a larger field of view, and an improved image quality. When finished, it will be possible to measure the star, shell, and background stars in the same exposures, reducing the error even further.

Besides R~Scl, there are six additional detached shell sources known~\citep{schoieretal2005,ramstedtetal2011,maerckeretal2014}, with shell radii ranging between $\approx$7\arcsec and 60\arcsec. This method could be applied to these sources as well. 

If the new PolCor instrument meets the expected improved image quality, it may also be possible to use this method to determine the distance to regular dusty envelopes around AGB stars. Also here the polarised observations give the distribution of the dust in the plane of the sky, reducing any confusion due to geometrical effects. As long as a dusty feature can be clearly identified, the variation in intensity compared to the stellar pulsations will give the absolute distance of the dusty feature to the star, and hence the distance to the star from Earth. The results for R~Scl indicate that, together with SED fitting of atmospheric models, phase lag measurements in dust-scattered stellar light might provide one of the most reliable distance estimates to individual AGB stars.

\begin{acknowledgements}
M.Maercker acknowledges support from the Swedish Research Council. M.B. acknowledges funding through the uni:docs fellowship of the University of Vienna, and funding by the Austrian Science Fund FWF under project number P23586. E.D.B. acknowledges funding by the Swedish National Spaceboard.
\end{acknowledgements}

\bibliographystyle{aa} 
\bibliography{../../../bib/maercker}

\end{document}